\documentclass[final,1p,times]{elsarticle} 
\usepackage{graphicx} 
\usepackage{amssymb} 
\usepackage{amsthm} 
\usepackage{lineno} 

\journal{Nuclear Physics A} 
\begin{document} 

\begin{frontmatter} 



\title{Study of the Fluctuations of Net-charge and Net-protons Using Higher Order Moments}

\author{Tapan K. Nayak$^{a}$ for the STAR collaboration}

\address[a]{Variable Energy Cyclotron Centre, 1/AF Bidhan Nagar, Kolkata - 700064, India} 

\begin{abstract} 

We present the STAR preliminary results on mid-rapidity and low transverse momentum
mean, standard deviation, skewness, and kurtosis of
net-charge and net-proton distributions in Au+Au and Cu+Cu collisions 
at $\sqrt{s_{NN}}$ = 200~GeV for various collision centralities. 
All the measured high moments of these distributions can be scaled by
the number of participating nucleons, consistent with the soft process emissions.
The ratios of fourth to second order cumulants of both the net-charge and net-proton 
distributions are consistent with models without QCD critical point. 

\end{abstract} 

\end{frontmatter} 



\section{Introduction}

One of the goals of the physics program at the Relativistic Heavy-Ion Collider (RHIC) is to locate
a critical point in the QCD phase diagram \cite{rajagopal}.
The QCD phase diagram is often plotted as temperature ($T$) {\it vs.} baryon chemical potential 
($\mu_{\rm B}$). Lattice Gauge Theory (LGT) calculations indicate that at small chemical 
potential ($\mu_{\rm B}$$\sim $0), corresponding to the cases for RHIC and LHC, the transition from hadronic
to partonic matter is a cross-over \cite{bernard}. On the other hand, model calculations suggest that
when the $\mu_{\rm B}$ is larger, 
the transition becomes first order in nature \cite{Scavenius:2000qd}. Therefore, 
one expects the existence of a critical point at the end of first
order transition \cite{rajagopal}. At RHIC we plan to scan \cite{bes} the center of mass energy
($\sqrt{s_{NN}}$) to vary $\mu_{\rm B}$ in order to search for the location of the
QCD critical point. Several lattice calculations suggest the existence of the critical point for 
$\mu_{B}$ $>$~160 MeV \cite{gupta,latticeqcp}, corresponding to a center of mass energy of
$\sqrt{s_{NN}}<20$~GeV. 

One of the characteristic signatures of the
critical point is an increase in fluctuations of various event-by-event 
observables~\cite{stephanov,koch}. 
The moments of the conserved quantities such as net-charge,
net-baryon and net strangeness are related to respective thermodynamic quantity susceptibilities. The LGT
calculations, indeed, have shown that the fourth moment of these distributions have large values or 
diverge at critical temperature near the QCD critical point ~\cite{gupta,gupta2,ejiri,lattice-cheng}.
It has been proposed that higher moments of 
event-by-event net-charge and net-proton multiplicities
are significantly 
more sensitive to existence of critical point compared to measures based 
on second moments~\cite{stephanov3}. The fourth moments of these multiplicity distributions 
are expected to be proportional to the seventh power of the correlation length~\cite{stephanov3}. 
It is expected that the evolution of fluctuations from the critical point 
to the freeze-out may lead to non-Gaussianity in the event-by-event 
multiplicity distributions.  Thus, the kurtosis of multiplicity 
distributions (net-charge or net-proton) could be a sensitive observable in
the search of QCD critical point.  

We present the results of the moments of
net-charge and net-proton distributions at midrapidity 
from Au+Au and Cu+Cu collisions at $\sqrt{s_{NN}}$ = 200~GeV in the STAR
experiment at RHIC.


\section{Analysis Details, Results and Discussions}

With the large acceptance and excellent particle identification capabilities, 
STAR provides the best environment for studying event by event fluctuations.
The charged particles were detected in the Time Projection Chamber (TPC) and
protons were identified by their specific energy loss in the TPC gas.
Standard STAR track quality cuts were applied in this analysis. The charged 
particle tracks are chosen to be within the transverse momentum ($p_{T}$) range of 
$0.15 - 1.0$~GeV/c whereas for protons the $p_{T}$ range is taken to be $0.4 - 0.8$ GeV/c
in order to have a clean proton identification and to reject secondary
protons from interactions with detector material. 
The analysis is carried out for Au+Au and Cu+Cu collisions at 
$\sqrt{s_{NN}}$ = 200~GeV for various collision centralities.

\begin{figure}[ht]
\centering
\includegraphics[scale=0.43]{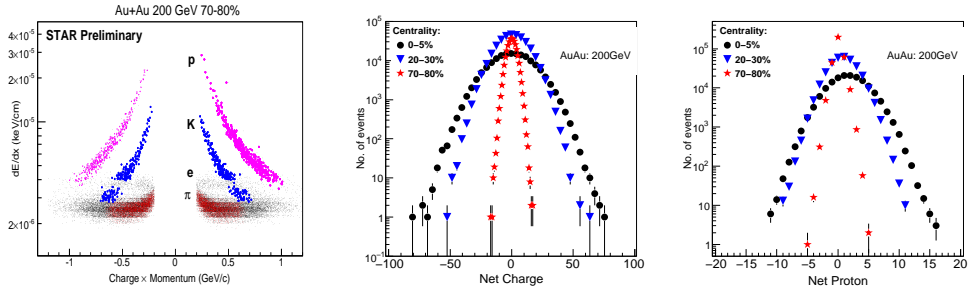}
\caption[]{
Left panel shows the $dE/dx$ (specific ionization loss) {\it vs}. rigidity distribution of charged particles
for 70-80\% most central bin in Au+Au collisions at $\sqrt{s_{NN}}$ = 200~GeV. 
Net-charge (middle panel) and net-proton (right panel) distributions for
Au+Au collisions at $\sqrt{s_{NN}}$ = 200~GeV for three different centralities.
The charged particle measurements are for $|\eta|<0.5$ and $0.15<p_T<1.0$~GeV/c wheres
the protons are within $|y|<0.5$ and $0.4<p_T<0.8$~GeV/c.
}
\label{Fig1}
\end{figure}

The left panel of Fig.~\ref{Fig1} shows a typical $dE/dx$ {\it vs.} rigidity (defined as
the momentum $\times$ the charge of the particle)
distribution for charged particles as measured in the STAR TPC for 70-80\% most central bin in
Au+Au collisions
at $\sqrt{s_{NN}}$ = 200~GeV. The figure gives a clean proton identification
for $p_{T}$ $<$ 1~GeV/$c$.
The net-charge ($h^{+}-h^{-}$)
and net-proton ($p-\overline{p}$) 
distributions for Au+Au collisions at $\sqrt{s_{NN}}$ = 200~GeV for three different
centralities (0-5\%, 20-30\% and 70-80\%) are also shown in Fig.~\ref{Fig1}. 
For the charged
particles the distributions are symmetric around the mean value whereas the
distributions for net-protons seem to be skewed in peripheral collisions. These distributions are
further analyzed in order to extract various moments.

The characterization of a given data set may be obtained by analyzing the moments
of their distribution.
For a given distribution of $N$ data points, the third and fourth moments are expressed in terms of
skewness and kurtosis as given by:
\begin{equation}
{\rm Skewness~~ = ~~} \frac{\langle (\delta N)^3 \rangle}{\sigma^3} {\rm ~~~~and~~~~~~~~}
{\rm Kurtosis~~ = ~~} \frac{\langle (\delta N)^4 \rangle}{\sigma^4} - 3,
\end{equation}
where $\delta N = N - \langle N\rangle$ and $\sigma = \sqrt{\langle (\delta N)^2 \rangle}$ is the standard deviation.
Skewness is a measure of the symmetry of a distribution whereas kurtosis measures the peakedness of
the distribution.
For a Gaussian distribution, the values of skewness and kurtosis are zero.
A positive value of skewness means a long tail on the right side
and a negative value means the left tail is dominant. 
A data set with positive value of
kurtosis is peaked at the center whereas a negative
kurtosis is obtained for a flatter top near the mean.

\begin{figure}[ht]
\centering
\includegraphics[scale=0.65]{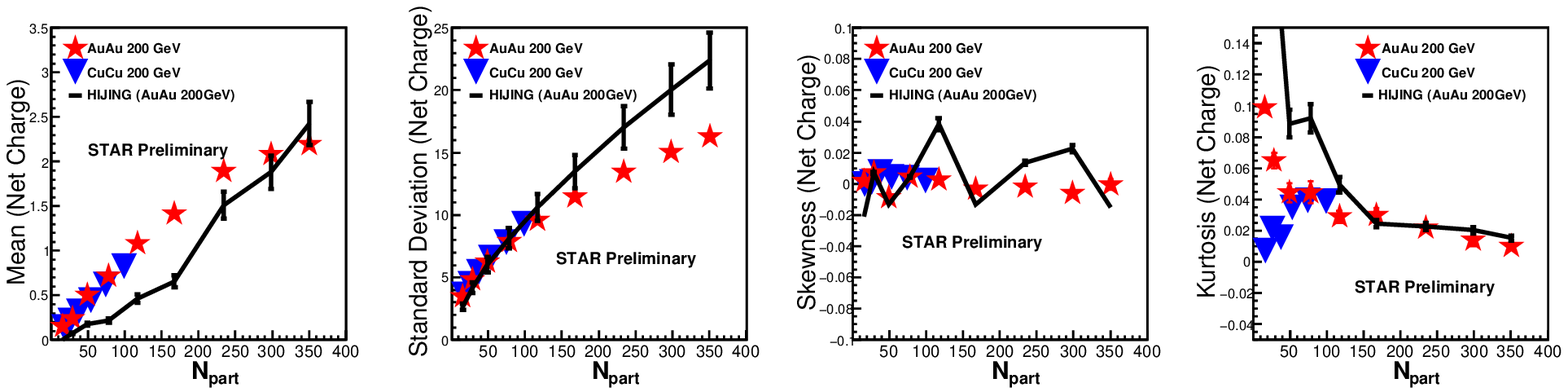}	       
\includegraphics[scale=0.65]{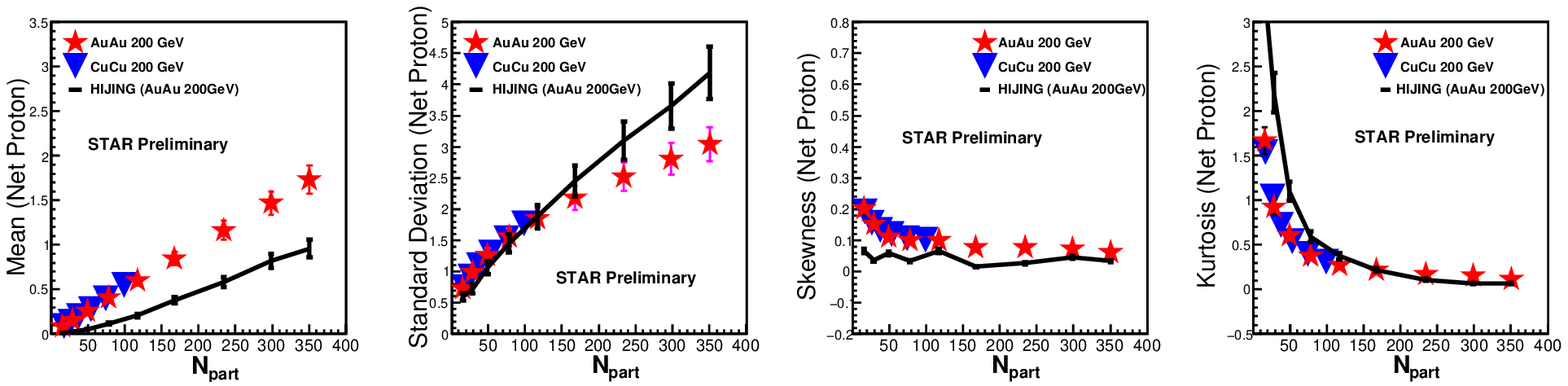}	       
\caption[]{Mean, standard deviation, skewness and kurtosis of 
net-charge (top panels) and net-proton (bottom panels) 
distributions as a function of centrality for $\sqrt{s_{NN}}$ = 200~GeV
Au+Au and Cu+Cu collisions. HIJING results for Au+Au 200~GeV collision
are also shown.}
\label{Fig2}
\end{figure}

Fig.~\ref{Fig2} shows the evolution of various moments of the net-charge 
and net-proton as a function of collision centralities (denoted
by the number of participant nucleons, $N_{\mathrm {part}}$) for Au+Au and Cu+Cu 
collisions at $\sqrt{s_{NN}}$ = 200~GeV. Note that the distributions have not
been corrected for reconstruction efficiency and acceptance. The error bars 
include both statistical and systematic errors obtained by varying
track quality conditions. In going from peripheral to central collisions 
we observe that the mean and standard deviation 
increase smoothly for all cases. The 
skewness of net-charge distributions are close to zero and similar for 
all centralities, whereas for net-protons, the skewness values have a 
decreasing trend in going from peripheral to central collisions. For 
net-charge distributions, the kurtosis values are small and has a decreasing 
trend in going from peripheral to central collisions for Au+Au systems. 
For net-protons, the higher values of kurtosis for peripheral collisions 
compared to central collisions indicate the distributions are more peaked 
at the mean value for peripheral collisions~\cite{bedanga_qm09}. The results from HIJING event 
generator for Au+Au at 200~GeV are shown by solid lines, which are qualitatively
similar to those of the data.

Several theoretical calculations based on Lattice QCD~\cite{gupta2,ejiri,lattice-cheng} and 
other QCD models~\cite{redlich} have suggested that the ratio of quartic to quadratic fluctuations of
the net quark number, defined as $R_{4,2}$ to be a probe of QCD critical point. It can be shown that:
\begin{equation}
R_{4,2}= \frac{\langle (\delta N)^4 \rangle}{\langle (\delta N)^2 \rangle} - 3{\langle (\delta N)^2 \rangle} = 
{\rm Kurtosis} \times {\rm Variance}. 
\end{equation}
$R_{4,2}$ eliminates the 1/N behaviour necessary for the central limit theorem and removes explicit dependence
on centrality. The significance of $R_{4,2}$ is that it is entirely dynamic and can be sensitive to the location of the
critical point~\cite{gupta2,ejiri,lattice-cheng,redlich}. 

\begin{figure}[ht]
\centering
\includegraphics[scale=0.5]{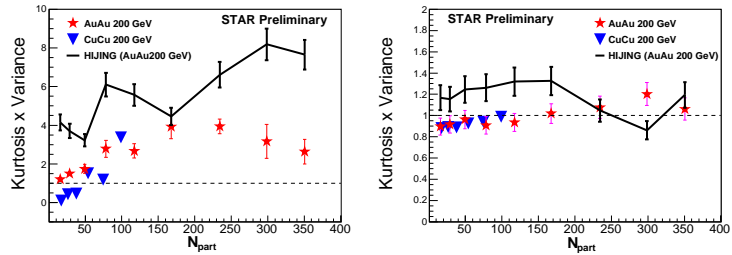}	       
\caption[]{$R_{4,2}$(={\it ~kurtosis~}$\times${\it ~variance}) of net-charge (left panel) and
net-proton (right panel) distributions for $\sqrt{s_{NN}}$ = 200~GeV Au+Au and 
Cu+Cu collisions. Results from HIJING are shown as solid lines. }
\label{Fig4}
\end{figure}

Fig.~\ref{Fig4} shows the $R_{4,2}$ values for net-charge and net-proton distributions. While the net proton values
are close to unity, net-charge values are higher. The results are similar to those obtained from HIJING.
While the HIJING event generator is not modelled on the
existence of QCD critical point, its ability to reproduce the
qualitative features of the data needs further investigation. 

In summary, we have presented a study of mean, standard deviation, skewness 
and kurtosis of event-by-event net-charge and net-proton distributions 
in $\sqrt{s_{NN}}$  = 200~GeV Au+Au and Cu+Cu collisions from the STAR 
experiment at RHIC. The results are presented as a function of collision
centrality  for
mid-rapidity and low $p_T$ ranges. In case of net-charge distributions, 
the skewness values are close to zero for all centralities indicating the 
distributions are rather symmetric around the mean and the $R_{4,2}$ 
distributions are consistent with the 
predictions from HIJING. For net-protons, the kurtosis distributions are 
observed to decrease with increasing collision centrality and the $R_{4,2}$ is
unity for all centrality bins. The values are in agreement with model 
calculations from HIJING which do not have the physics of QCD 
critical point. Detailed comparison with other event generators, such as
UrQMD, is in progress~\cite{schuster}.

The present study at the currently available RHIC energies only probes 
the $\mu_{B}$ $<$ 30 MeV region of the phase diagram, while most theoretical 
calculations expect the QCD critical point to exist around $\mu_{B}>$160~MeV. 
The current study will help to understand the expectations from various  
physics processes to the forthcoming RHIC critical point search 
program~\cite{bes}. 

\section*{Acknowledgments}
We thank Dr. S. Gupta, F. Karsch, V. Koch, K. Rajagopal, K. Redlich and 
M. Stephanov for exciting discussions on the subject.



\begin{thebibliography}{00} 
   
\bibitem{rajagopal} K. Rajagopal and F. Wilczek, The Condensed matter physics of QCD, arXiv:hep-ph/0011333.

\bibitem{bernard} C. Bernard {\it et al.}, Phys. Rev. D {\bf 75}, 094505 (2007).

\bibitem{Scavenius:2000qd}
   O.~Scavenius, A.~Mocsy, I.~N.~Mishustin and D.~H.~Rischke,
   Phys.\ Rev.\  C {\bf 64}, 045202 (2001).

\bibitem{bes} B. I. Abelev {\it et al.}, STAR Collaboration, STAR Internal Note - SN0493.   

\bibitem{gupta} R. Gavai and S. Gupta Phys. Rev. D {\bf 78}, 114503 (2008).

\bibitem{latticeqcp} Z. Fodor and S.D. Katz JHEP {\bf 0404}, 50 (2004).

\bibitem{stephanov} M. A. Stephanov {\it et al.}, PRD {\bf 60}, 114028 (1999).

\bibitem{koch} V. Koch, arXiv:0810.2520 [nucl-th].

\bibitem{gupta2} R. Gavai and S. Gupta, Phys. Rev. D {\bf 71}, 114014 (2005).

\bibitem{ejiri} S. Ejiri {\it et al.}, Nucl. Phys. Proc. Suppl. {\bf 140} 505 (2005).



\bibitem{lattice-cheng} M. Cheng {\it et al.}, arXiv:0811.1006 [hep-lat].

\bibitem{stephanov3} M. A. Stephanov, Phys. Rev. Lett. {\bf 102} (2009) 032301.

\bibitem{bedanga_qm09} B. Mohanty {\it et al.}, STAR Collaboration, Poster presented at Quark Matter 2009.

\bibitem{redlich} Stokic, Friman and Redlich, Phys. Lett. {\bf B673}, 192 (2009).

\bibitem{schuster} T. Schuster {\it et al.}, arXiv:0903.2911 [hep-ph]. 

\end{thebibliography}
\end{document}